\documentclass[fleqn,usenatbib]{mnras}

\usepackage[utf8]{inputenc}

\usepackage{acronym}
\usepackage{ae,aecompl}
\usepackage{booktabs}
\usepackage{dcolumn}
\usepackage{graphicx}
\usepackage{amsmath,amsfonts,amssymb}
\usepackage{mathrsfs}
\usepackage[normalem]{ulem}
\usepackage{epstopdf}

\usepackage[T1]{fontenc}

\let\oldhref\href
\renewcommand{\href}[2]{\oldhref{#1}{\hbox{#2}}}

 \usepackage{soul,xcolor}
 \setstcolor{red}

\title[Resolution Dependence of CCSN simulations]{Towards an Understanding of the Resolution 
Dependence of Core-Collapse Supernova Simulations}

\author[H. Nagakura et al.]{
Hiroki Nagakura$^{1}$\thanks{E-mail: hirokin@astro.princeton.edu},
Adam Burrows $^{1}$,
David Radice$^{1,2}$,
David Vartanyan$^{1}$
\\
$^{1}$Department of Astrophysical Sciences, Princeton, NJ 08544, USA\\
$^{2}$Institute for Advanced Study, 1 Einstein Drive, Princeton, NJ 08540, USA\\
}

\date{Accepted XXX. Received YYY; in original form ZZZ}

\pubyear{2019}

\begin{document}
\label{firstpage}
\pagerange{\pageref{firstpage}--\pageref{lastpage}}
\maketitle

\begin{abstract}
Using our new state-of-the-art core-collapse supernova (CCSN) code F{\sc{ornax}}, we explore the 
dependence upon spatial resolution of the outcome and character of three-dimensional (3D) supernova simulations.
For the same 19-M$_{\odot}$ progenitor star, energy and radial binning, neutrino microphysics, and nuclear 
equation of state, changing only the number of angular bins in the $\theta$ and $\phi$ directions, 
we witness that our lowest resolution 3D simulation does not explode. However, when jumping progressively up 
in resolution by factors of two in each angular direction on our spherical-polar grid, models then explode, 
and explode slightly more vigorously with increasing resolution. This suggests that there can 
be a qualitative dependence of the outcome of 3D CCSN simulations upon spatial resolution. 
The critical aspect of higher spatial resolution is the adequate capturing of the physics
of neutrino-driven turbulence, in particular its Reynolds stress.  The greater numerical 
viscosity of lower-resolution simulations results in greater drag on the turbulent 
eddies that embody turbulent stress, and, hence, in a diminution of their vigor. Turbulent stress 
not only pushes the temporarily stalled shock further out, but bootstraps a concomitant 
increase in the deposited neutrino power. Both effects together lie at the core of the resolution 
dependence we observe. 
\end{abstract}

\begin{keywords}
starts - supernova - general
\end{keywords}

\section{Introduction}\label{sec:intro}

The theoretical viability of a three-dimensional (3D) model of a core-collapse supernova (CCSN)
explosion is contingent upon the fidelity of its microphysical and numerical representations.
The former involves the nuclear equation of state  
\citep{1991NuPhA.535..331L,2009ApJ...707.1495S,2013ApJ...774...17S,2017ApJ...848..105T,2017JPhG...44i4001F,2017NuPhA.961...78T,2017arXiv170701527D,2017RvMP...89a5007O}
and the neutrino opacities and emissivities \citep{2006NuPhA.777..356B}.
The latter involves not only the coupled equations to be solved, but the numerical algorithms
to solve them and the discretizations in space and neutrino momentum space (collectively phase space) 
employed.  The achievable density in phase space and the associated zoning are limited by the available 
computational resources and by the alacrity with which a particular code can step through time
as the dynamical evolution proceeds.  In all these regards, it is gratifying to report that the 
last decades have witnessed significant improvements on all fronts.  The ambiguities in 
the microphysics now seem modest and managable, though many-body corrections to the 
neutrino-matter interaction \citep{1998PhRvC..58..554B,sawyer1999,roberts2012,2017PhRvC..95b5801H,roberts_reddy2017,2018SSRv..214...33B}
and aspects of the nuclear equation of state are still evolving.  
Numerical techniques for solving the coupled equations of radiation/hydrodynamics
in the 3D context have taken great strides, though long-term simulations with multi-angle transport
are still a challenge \citep{2012ApJS..199...17S,2014ApJS..214...16N,2018ApJ...854..136N}.  Nevertheless, 3D simulations 
that incorporate all physical processes to an acceptable degree of accuracy have recently 
emerged \citep{2015ApJ...807L..31L,bmuller_2015,melson:15b,2016MNRAS.461L.112T, 2016ApJS..222...20K,
muller2017,2019ApJ...873...45G,2018ApJ...865...81O,summa2018,2019MNRAS.482..351V,2019MNRAS.485.3153B,2019arXiv190408088N}.  
Importantly, some groups can now generate many 3D simulations per year and can thereby perform comprehensive, in-depth 
investigations of the dependence of explosion upon physical parameters and numerical setup.
In contrast to years passed, most default multi-dimensional models now explode 
without artifice.  This includes detailed 3D models.

Whether a particular theoretical 3D model explodes is always in the context of a 
choice of 
spatial zoning.
Compromises are made to ensure
the simulation renders a significant result within resource constraints. However, the accuracy
with which a simulation captures Nature is not always clear.  This will depend not only upon
the physics and equations incorporated into the code, but upon the chosen resolutions. Numerical
dissipation and inaccuracies in a discritization scheme due to the chosen stencil, approximations 
to the derivatives and fluxes embedded in the equations, and the method of time-stepping 
can compromise the results.  In the recent past, we have explored the dependence of our 
2D simulations upon the number of neutrino energy groups and to date have found surprisingly little variation.  
In the future, we plan to communicate the results of these studies and to update them to include 3D 
models, but more should be done in this regard. 

However, spatial resolution studies in the context of state-of-the-art 3D CCSN 
codes are rare.  Previous such studies in the context of CCSNe
 mostly focused on the resolution of the turbulence behind the stalled 
shock wave and the associated requirements to resolve the full inertial range
\citep{2010PhRvD..82j3016N,2014ApJ...785..123C,2015ApJ...808...70A,2015ComAC...2....7R,2016ApJ...820...76R}.
In particular, \citet{2016ApJ...820...76R}, in a transparently straightforward
and systematic fashion, scrutinized the kinetic-energy power spectrum and the
approach to the Kolmogorov cascade and explored the effective numerical
viscosities as a function of resolution. They concluded that an inertial
range appeared only in their highest angular resolution ($\sim$0.09$^{\circ}$).
However, their study used a variant of a simple ``light-bulb" neutrino heating
scheme that did not include crucial physical effects such as the feedback
between accretion rate and neutrino luminosity.  \citet{2015ApJ...799....5C} did look
not only at the effect of resolution on the character of turbulence, but also on explodability,
and found a connection between resolution and susceptibility to explosion. 
However, they too used a simple neutrino leakage scheme. \citet{2016ApJ...831...98R}
did perform collapse simulations with a viable transport algorithm that
consistently coupled with the hydrodynamics. However, they did not include
inelastic scattering and neutrino energy redistribution, nor the velocity
dependence of the transport. They performed calculations at low and high 
resolution, and for both a single octant (solid angle $\pi/2$) and the full $4\pi$ 
steradians, for a total of four simulations. They found that, contrary to our findings,
low resolution was more explodable and that calculating in a single octant
slightly inhibited explosion. The reasons for the difference between what we report here 
and that work may lie with our more complete physics suite, differences in the grids,
or differences in the progenitors, but remains unresolved. 
\citet{2012ApJ...749...98T} focused on exploding models and 
conducted two 3D simulations of explosion at different $\phi$ resolutions, employing the
reduced dimension ``ray-by-ray" transport approach with the IDSA method. They
concluded that, though both models exploded, the higher resolution model exploded
more vigorously. However, the grid was comprised of only 64 $\theta$ angular bins and
either 32 (their low-resolution) or 64 $\phi$ angular bins and only 300
logarithmically-spaced radial zones. The binning is rather coarse and may not 
capture important characteristics of the turbulence. Moreover, these calculations did not
include $\nu_{\mu}$ and $\nu_{\tau}$ neutrinos nor
energy redistribution by inelastic scattering.

Recently, \citet{2019arXiv190401699M} did look at the question of the resolution
dependence of explodability, but their most solid conclusions involved a series of
3D calculations with light-bulb heating that were not consistent with their hydrodynamics.
The highest angular resolution achieved in those approximate 3D simulations
was 0.5$^{\circ}$ and they used an implementation of a Yin-Yang grid
\citep{2004GGG.....5.9005K,2010A&A...514A..48W} to avoid the axial coordinate singularity
at the poles of their otherwise spherical-polar ($r\times\theta\times\phi$) coordinate
system. In the inner 42$-$46 km, these light-bulb simulations were done in spherical
symmetry (1D), thereby suppressing PNS convection. As expected, in those simulations
\citet{2019arXiv190401699M} witnessed progressively better resolved turbulence. They also
witnessed a more vigorous explosion with improving angular resolution. This is in contrast
with the allied results of \citet{2012ApJ...755..138H}, who suggested with their light-bulb
study that increasing resolution actually inhibited explosion. \citet{2019arXiv190401699M}
speculated
 that the reason for the different trends they found in 2012 and in 2019
resided
in the possible resolution dependence of the numerical seed perturbations imposed when
using spherical-polar coordinates without the Yin-Yang grid.  However, as we show in this paper,
we see no such artifact and witness in our suite of three 3D full-physics spherical-polar
simulations increasing explodability with increasing angular resolution, all else being exactly
the same. In fact, our lowest resolution model does not explode, while the two highest resolution
models do.  This is a reassuring trend that suggests increasing the resolution even further,
in a way Nature does effortlessly, will not in itself compromise the conclusions concerning
explodability, and that pushing to further resolve the full Kolmogorov inertial range will not change
our 3D simulation results qualitatively. \citet{2019arXiv190401699M} did perform four
fully-consistent radiation-hydrodynamic calculations. They studied a $9M_\odot$ progenitor that
exploded nearly identically in a low resolution setup with 3.5$^{\circ}$ angular bins and with
a static-mesh-refinement (SMR) simulation that achieved an angular resolution as high as
0.5$^{\circ}$.  This model had been shown recently to explode easily and early
\citep{2019MNRAS.485.3153B,2019ApJ...873...45G}. They also compared the evolution of a $20M_{\odot}$ progenitor at low-resolution
(2$^{\circ}$) using a spherical grid, and at higher resolution using their implementation of
SMR and a Ying-Yang grid.  However, their higher-resolution SMR run did not explode, while
their lower resolution run did, evincing thereby the opposite trend to their light-bulb study.
They speculated that their implementation of SMR, through it conserved total thermal
plus kinetic energy by construction, artificially converted an excess of kinetic energy into
thermal energy when matter traversed the refinement boundaries, at which their implementation
was a 2-to-1 (de)refinement.  This would mute the turbulent pressure shown to be central to
explosion \citep{1995ApJ...450..830B,murphy:13,2017ApJ...850...43R,2019MNRAS.485.3153B},
since the effective gamma of turbulent energy is higher than that for gas ($\sim$2 vs.
$\sim$4/3) \citep{1995ApJ...450..830B,2016ApJ...820...76R}. However, we use a dendritic grid
that employs our implementation of SMR (described in detail in \citet{2019ApJS..241....7S}) and
do not see this effect. The 3D calculations we present here incorporate the full suite of
necessary physics, are performed with the state-of-the-art 3D radiation-hydro code F{\sc{ornax}}
\citep{2019ApJS..241....7S,2019MNRAS.485.3153B,2018MNRAS.477.3091V,2019MNRAS.482..351V}, and are done
changing only the angular resolution, keeping all else exactly the same. Ours is therefore one of the
first resolution studies of the 3D radiation/hydrodynamics of the neutrino-driven explosion mechanism 
incorporating all the physical effects though important and addressing, in the main, 
the full computational challenge.

We first describe in \S\ref{sec:methodandmodel} our F{\sc{ornax}} CCSN code and model setup.
Then in \S\ref{results}, we present our results.  This section includes a discussion of 
the resolution dependence of the dynamics, neutrino emissions and heating, turbulence, 
and turbulent stress. As stated above, we find that as the angular/spatial resolution is increased
the 3D model is more ``explodable," with the lowest resolution model not exploding at all. 
This suggests a stark, at times qualitative, dependence upon resolution.  This also suggests, but 
does not prove, that published non-exploding 3D models (e.g., the 13-M$_{\odot}$ model in 
\citet{2019MNRAS.485.3153B}) might explode at still higher resolution. In section \S\ref{sec:conclusions}, 
we summarize our results and speculate on their import.

\section{Model and Method} 
\label{sec:methodandmodel} 

\begin{figure*}
  \begin{minipage}{0.9\hsize}
	\includegraphics[width=\columnwidth]{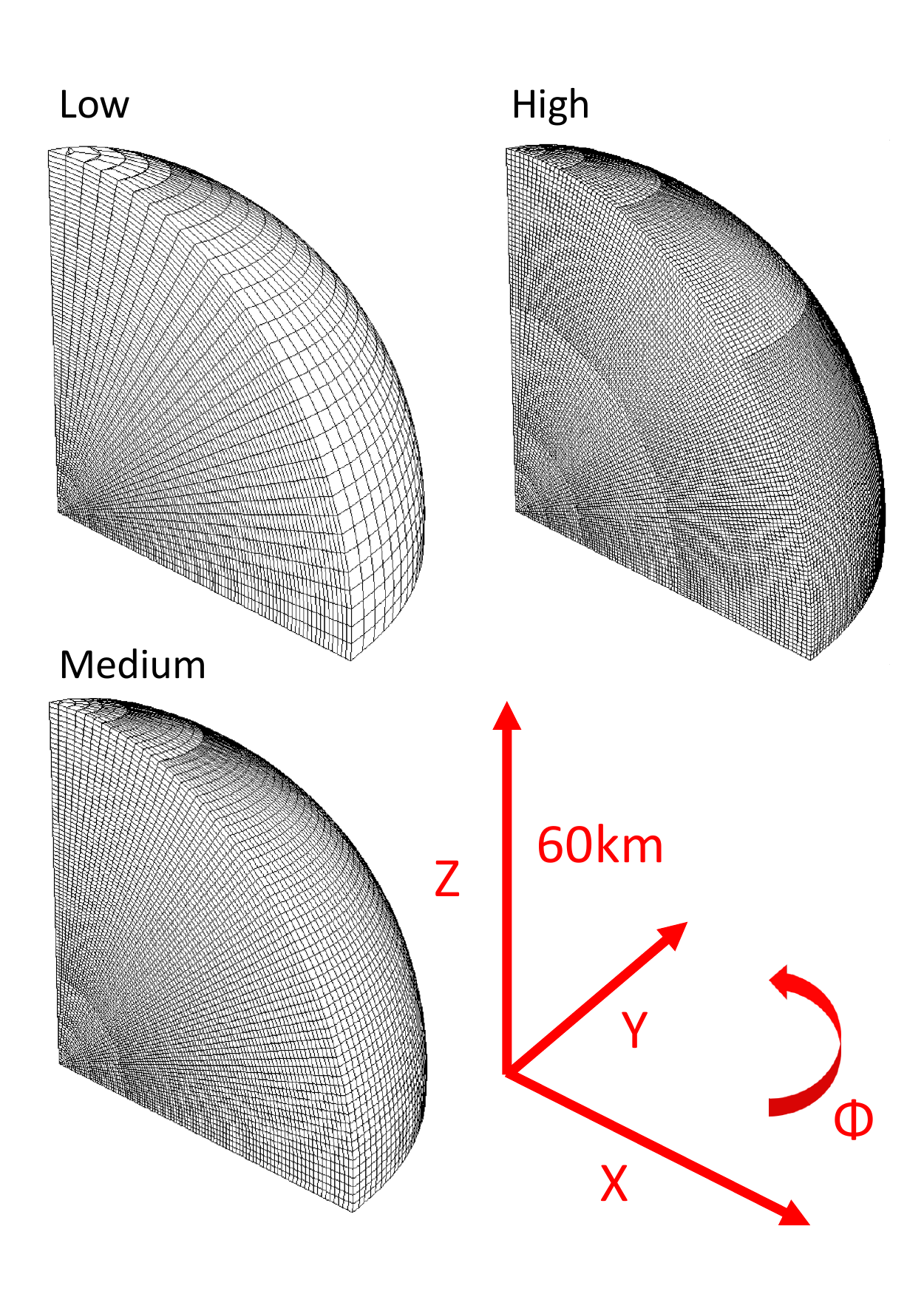}
    \caption{Octant cutouts portraying the static mesh setup in the inner 60 kilometers
for the three resolutions employed in our study.  Note that the number of refinement 
levels increases with increasing resolution, but that the dendritic grid in $\theta$
does not extend for any model beyond $\sim$35 kilometers.  However, refinement in $\phi$ 
is performed over the entire grid to avoid an otherwise challenging CFL constraint 
along the polar axis. All models have the same radial zoning.}
    \label{Gridstructure_vertical_v2}
  \end{minipage}
\end{figure*}

To explore the dependence of the character and outcome of 3D
core collapse upon the spatial resolution of the computational grid, 
we focus in this paper on the 19 $M_\odot$ ZAMS mass progenitor model 
of \citet{2016ApJ...821...38S}.  Three spherical-polar models are simulated that are
equivalent in every way except for the number of angular zones employed.
The low resolution model (3DL) has 64 zones in $\theta$ and 128 zones in $\phi$,
the medium (default) resolution model (3DM) has 128 zones in $\theta$ and 
256 zones in $\phi$, and the high resolution model (3DH) has 256 zones in $\theta$ 
and 512 zones in $\phi$\footnote{When performing 2D simulations for comparison, 
we use the same $\theta$ resolution employed in the corresponding 3D model. 
Note that the 2D models were simulated merely for comparison purposes 
and are not important for this investigation.}.  
Therefore, these simulations follow 0.25, 1.0, and 4.0 times
as many zones as our default resolution runs, such as those published in 
\citet{2019MNRAS.482..351V,2019MNRAS.485.3153B}. All models encompass the full $4\pi$ steradians and 
are run with 678 zones in radius out to 20,000 kilometers (km), with a center 
radial zone width of 0.5 km, and a dendritic grid in $\theta$ out to $\sim$10 km (3DL), $\sim$15 km (3DM), and $\sim$35 km (3DH), 
exterior to which the radial zoning is logarithmic. The $\phi$ mesh refinement to tame the
vertical axis is conducted on the entire grid. The dendritic grid employed in F{\sc{ornax}} is 
equivalent to static-mesh refinement and is described in detail in \citet{2019ApJS..241....7S}.
We use it to eliminate the otherwise severe Courant (CFL) timestep limits in the angular 
directions near the center and on the vertical axis of a spherical-polar grid.  The result 
is a code whose timestep is limited solely by the CFL condition in the radial zones
and is thereby more than $\sim{5}\times$ faster than conventional spherical-polar 
codes that would encompass the entire core.  Typical timesteps after bounce are 
$\sim$1 microsecond. 
The refinement is 2-to-1 and is done to very approximately maintain the aspect ratio of the grid cubes.
This allows us to simulate 
to the very center and to follow proto-neutron star (PNS) convection without inhibition.  
All other supernova codes with spherical-polar grids perform the inner simulations 
in 1D (or as an imposed boundary), even when the outer grid is either 2D or 3D. 
Moreover, our low-resolution model (3DL) boasts a resolution similar to (and 
often better than) the standard resolutions found in the papers of others 
employing a spherical-polar grid \citep{2012ApJ...749...98T,2014ApJ...792...96T,2014ApJ...786...83T,
bmuller_2015,2016ApJS..222...20K,muller2017,2016MNRAS.461L.112T,2018MNRAS.475L..91T,summa2018,
2019ApJ...873...45G}.  Our highest resolution model (3DH) is one of the highest resolution 
calculations ever performed in 3D core-collapse theory using a spherical-polar grid.  
Figure \ref{Gridstructure_vertical_v2}  depicts the grid and refinement structure near and in the space tiled dendritically.   

To conserve resources, the infall and bounce phases for all the multi-D runs in 
this paper were performed in 1D (spherical) and then mapped to the 3D (or 2D) 
grid 10 ms after bounce.
In the mapping of the matter field from 1D 
to multi-D, we add a small (dynamically unimportant) perturbation to the radial velocity 
with a maximum amplitude of 100 km ${\rm s}^{-1}$ in the region $200 < r < 1000$ km, following 
the prescription of \citet{2015MNRAS.448.2141M} with $\ell = 4$ and $n = 10$. 
This seed perturbation method is further described in \citet{2017ApJ...850...43R}.

All simulations presented in this paper were performed using the new F{\sc{ornax}} code. 
F{\sc{ornax}} is a 1D, 2D, and 3D radiation/hydrodynamics code that incorporates
the full suite of physical processes thought to be of relevance in core collapse and explosion
and is described in detail in \citet{2019ApJS..241....7S}, \citet{2018MNRAS.477.3091V}, and \citet{2018SSRv..214...33B}.
To date, results using F{\sc{ornax}} have been published in numerous papers \citep{2016ApJ...817..182W,2016ApJ...831...81S,
2017ApJ...850...43R,2018SSRv..214...33B,2018MNRAS.477.3091V,2018ApJ...861...10M,2019MNRAS.482..351V,
2019MNRAS.485.3153B,2018arXiv181207703R} addressing many aspects of the CCSN phenomenon.
F{\sc{ornax}} employs a multi-group two-moment transport method with analytical M1 
closures for the second and third moments \citep{2011JQSRT.112.1323V}.  The vector neutrino 
flux (first-moment) is fully solved and we do not employ the problematic ``ray-by-ray+" dimensional
compromise in which multiple 1D radial transport solves that ignore transverse transport
take the place of a multi-dimensional solution.  The latter has been shown to introduce 
anomalous behavior when the dynamics are not approximately spherical \citep{2016ApJ...831...81S,2019ApJ...873...45G}.  
F{\sc{ornax}} solves the transport equations in the comoving-frame, includes velocity-dependent 
frequency advection to order $v/c$ and the gravitational redshift effect \citep{2011PThPh.125.1255S},
factors in inelastic neutrino-electron and neutrino-nucleon scattering energy redistribution using the 
method found in \citet{2003ApJ...592..434T}, \citet{burrows_thompson2004}, and \citet{2006NuPhA.777..356B},
and employs the neutrino-matter interactions detailed in \citet{2006NuPhA.777..356B}.  Weak magnetism
and recoil corrections to neutrino-nucleon scattering and absorption rates are incorporated
using the prescriptions of \citet{2002PhRvD..65d3001H} and the many-body correction to the 
axial-vector term in neutrino-nucleon scattering is included following \citet{2017PhRvC..95b5801H}.
We distinguish three neutrino species: electron-type ($\nu_{\rm e}$), electron anti-neutrino 
type ($\bar{\nu}_{\rm e}$), and all others collectively denoted ``$\nu_x$"s.  The twelve
energy groups range logarithmically from 1 to 100 MeV for the $\bar{\nu}_{\rm e}$s and $\nu_x$s and 
from 1 to 300 MeV for the $\nu_{\rm e}$s.  Due to the narrower energy range for the 
$\bar{\nu}_{\rm e}$s and $\nu_x$s, the density of energy bins for them is higher 
than traditionally employed in the literature, particularly in 3D simulations.

Our choice of different energy binning for the different species 
arises from the fact that, at the high densities in the inner core, electron
neutrino transport dominates energy transport and the Fermi level of degenerate 
electron neutrinos is high enough to require the energy grouping extend
to high enough values for them. Electron neutrino degeneracy in the core, 
however, suppresses electron anti-neutrino densities severely and so their 
energy grouping need extend to only 100 MeV, allow a more highly-resolved
energy grid. Since the heavy-neutrinos are not degenerate and are subdominant 
energy carriers in the core, we employed the same narrower and
more highly-resolved energy range for them.

The equations are solved using a directionally-unsplit Godunov-type finite-volume 
scheme with HLLC fluxes for the hydrodynamics and HLLE fluxes for the radiation \citep{2019ApJS..241....7S,2019MNRAS.482..351V}.
The reconstruction is accomplished via a novel algorithm we developed
specifically for Fornax that uses moments of the coordinates within each
cell and the volume-averaged states to reconstruct TVD-limited parabolic
profiles, while requiring one less "ghost cell" than the standard PPM
approach.  The profiles always respect the cells' volume averages and, in
smooth parts of the solution away from extrema, yield third-order accurate
states on the faces.  We have taken special care in treating the
reconstruction of quantities on the mesh and have formulated our
reconstruction methods to be agnostic to choices of coordinates and mesh
equidistance.
Corrections for the enhanced gravity of general-relativity are handled using the now-standard 
``TOV-like" approach suggested by \citet{2006A&A...445..273M}.  Since in the core-collapse
problem the speed of light and the speed of sound are not that different, the spatial 
operators in the transport equations are solved explicitly.  This obviates the need
for complicated global iterative solvers, keeps the radiation solution local, and thereby
speeds up the code by factors of more than three. However, the local source terms, including
those due to inelasticity, are solved fully implicitly. For all models, we use the SFHo 
equation of state (EOS) \citep{2013ApJ...774...17S}, which is currently consistent 
with all known nuclear experimental and astrophysical constraints \citep{2017ApJ...848..105T}.

\section{Results} \label{results}

\subsection{Overall dynamics}
\label{sec:ovdy}

\begin{figure*}
  \begin{minipage}{0.75\hsize}
        \includegraphics[width=\columnwidth]{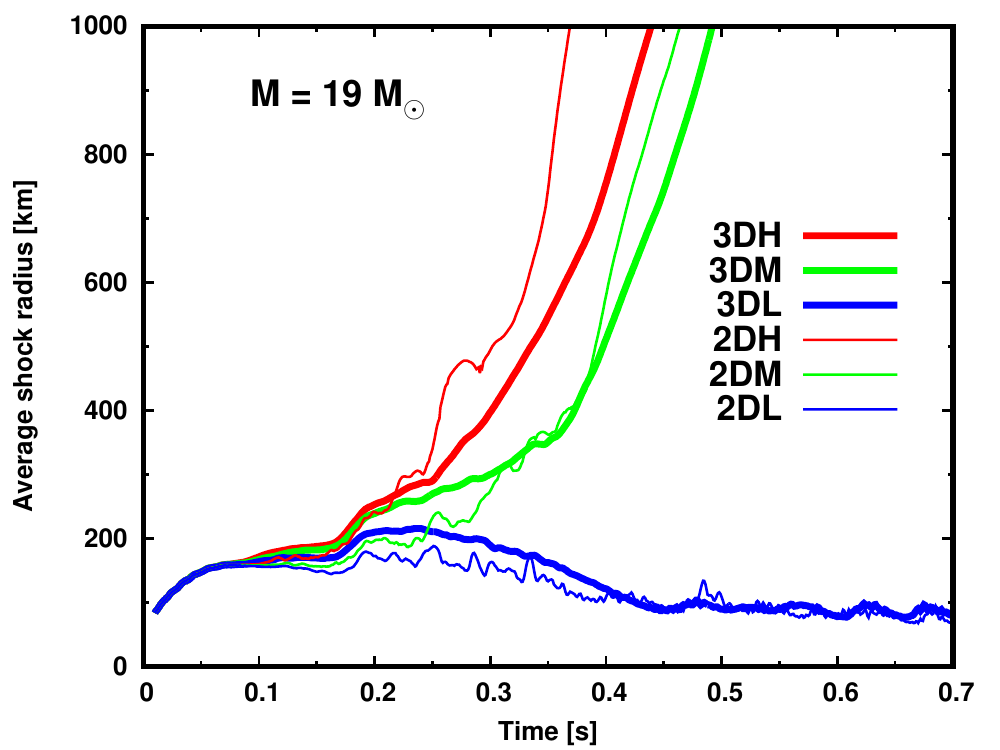}
    \caption{The mean shock radius (in km) versus time since bounce (in seconds) for
our 3D (thick) and 2D (thin) models.  We observe that models 3DL and 2DL (blue) do not explode, while 
models 3DM, 2DM, 3DH, and 2DH do explode.  All models use the same 19-M$_{\odot}$ progenitor 
from \protect\citet{2016ApJ...821...38S}.  Note that before $\sim$200 milliseconds the mean shock radius
is a slightly increasing function of resolution. See text for a discussion.}
    \label{graph_timetrajectories_shockradii}
  \end{minipage}
\end{figure*}
\begin{figure*}
  \begin{minipage}{1.0\hsize}
        \includegraphics[width=\columnwidth]{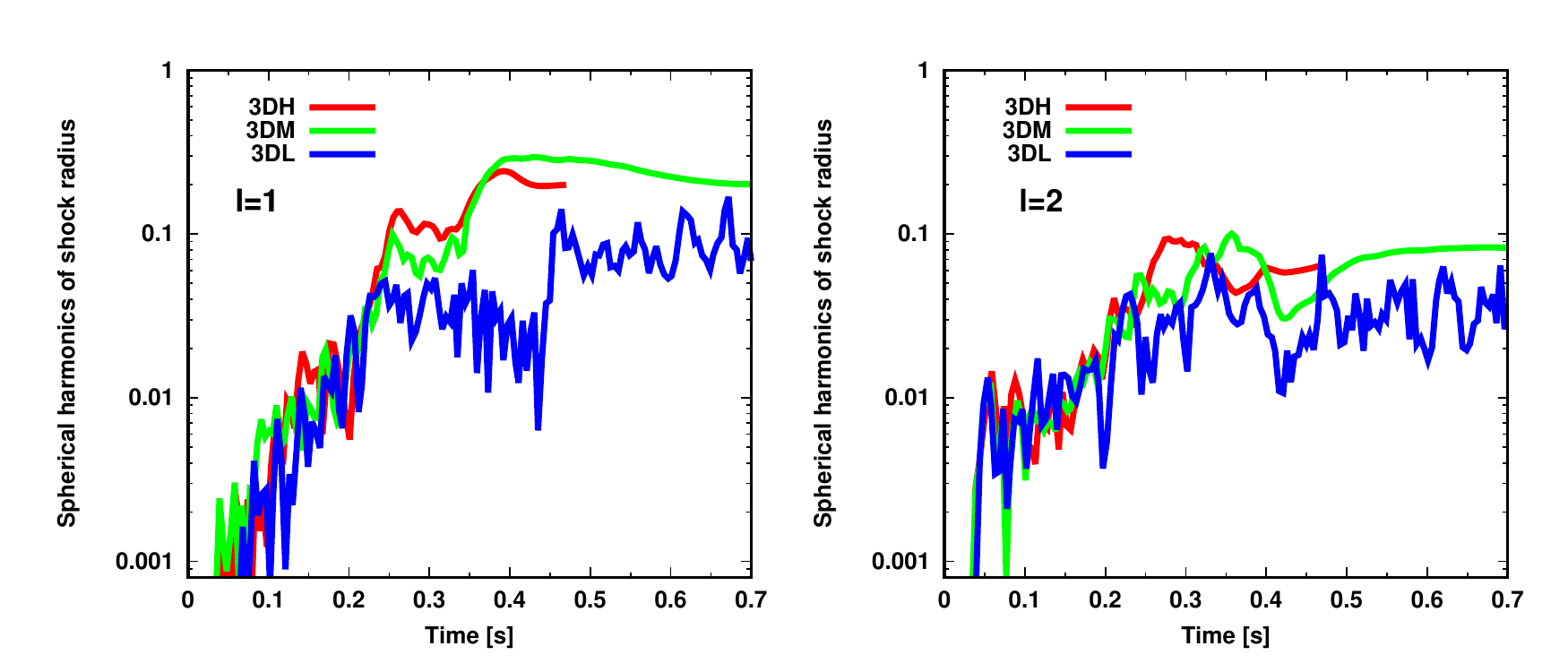}
    \caption{Renditions of the dipole ($\ell = 1$) and quadrupole ($\ell = 2$) components of the 
shock radius (normalized to the monopole term) versus time after bounce (in seconds).  The linear phase
of growth of these non-spherical distortions is followed by a non-linear phase in which the magnitude
of the associated distortion is a slightly increasing function of resolution.  Explosion near $\sim$200 ms
(when it occurs) is accompanied by manifest shock asphericities.}
    \label{graph_sphericalharmoRsh_s19}
  \end{minipage}
\end{figure*}

Figure \ref{graph_timetrajectories_shockradii} portrays the time evolution of the mean shock
radius after bounce of models 3DL, 3DM, and 3DH, as well as that of their 2D counterparts.
It demonstrates that whether this 19-M$_{\odot}$ progenitor model explodes depends upon the angular 
resolution chosen, with the lowest-resolution model 3DL not exploding and the two higher resolution
models exploding.  We note that before explosion between $\sim$100 ms and $\sim$200 ms the mean 
shock radii increase with resolution. We will return to this            
difference in \S\ref{turbulence}. The mean shock radius for the highest resolution model (3DH) launches more
quickly than that for the medium resolution model (3DM). In each case, the ``explodability" 
of the model is similar in 2D and 3D, with, however, the 3D models exploding a bit earlier and the 
2D models catching up a bit later.  Also, we continue to see that if a model explodes in 2D it generally explodes in 3D,
and vice versa for duds.  These patterns vis \`a vis 3D vs. 2D recapitulate what has been seen previously 
\citep{2019MNRAS.485.3153B,2019MNRAS.482..351V}\footnote{Since the neutrino signatures before explosion 
in 2D and 3D are similar, it is not surprising that there are crude similarities in outcome. 
In our estimation, part of the reason this has not been incorporated into the lore of supernova 
theory has been that most previous 2D simulations were done with the ``ray-by-ray" approach, which 
exaggerates the explodability of 2D models \citep{2016arXiv161105859B}.  Hence, in the literature 2D 
models that exploded were oftimes not accompanied by 3D models that did at or near the same times.  
Moreover, 3D runs were expensive, so the 3D models were not continued out very long.  The net 
appearance was a qualitative difference in outcome, one exploded while the other did not. If 
such models would have eventually exploded at later times, many modelers might have missed this. 
This is not to say 2D and 3D shouldn't be different (and we see this), if only because the character 
of the turbulence in 2D is different. Also, including as we do the effects of many-body suppression 
of neutrino-nucleon scattering, thereby making it slightly easier for both 2D and 3D models 
to explode \citep{2018SSRv..214...33B}.}.  Explosion for this progenitor
occurs early just as the turbulence behind the stalled shock wave achieves some degree of vigor.  One
measure of the growth of turbulent activity is provided in Figures \ref{graph_sphericalharmoRsh_s19},
which documents the increase in the amplitude of the dipolar and quadrupolar components of the shock 
radius with time after bounce.  These components have been normalized to the monopole, using 
the normalization convention of \citet{burrows2012}. Once the 
linear phase of the turbulent/Rayleigh-Taylor-like growth becomes non-linear 
(near time $\sim$200 ms), both the 3DM and 3DH models explode. This also 
approximately coincides with the accretion of the silicon/oxygen interface for this model
\footnote{We note that the mass accretion history for all models before explosion is exactly the same.}.  However,
the low-resolution 3DL model fails to launch and its dipolar shock oscillation saturates at a slightly 
lower value. This may not be coincidental (see \S\ref{turbulence}).

\begin{figure*}
  \begin{minipage}{0.7\hsize}
        \includegraphics[width=\columnwidth]{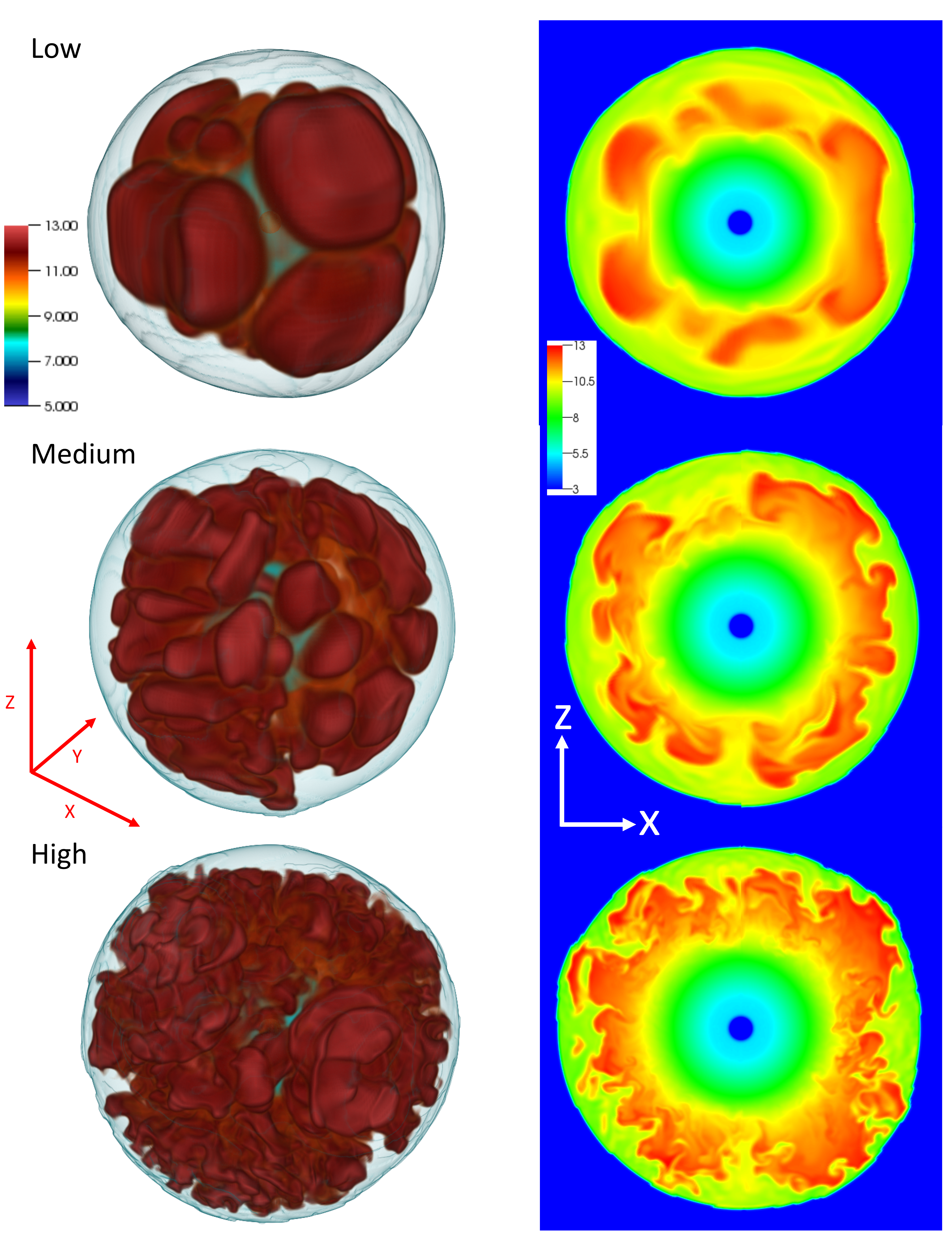}
    \caption{{\bf Left:} 3D volume renderings of the entropy of the core interior to the shock wave for
the three resolutions employed in this investigation at 100 ms after bounce.  The blue veil traces 
the shock wave. The scale of the box for each model is $2\times{190}$ km. {\bf Right:} The same regions shown as 
2D slices of the corresponding 3D models at the same time. The increasing facility with which the smaller structures 
are resolved as the grid resolution is improved is clearly brought out. We embed color bars and axes to clarify this mapping and the plot 
orientations, respectively.}
    \label{3D_ent1}
  \end{minipage}
\end{figure*}

\begin{figure*}
  \begin{minipage}{0.7\hsize}
        \includegraphics[width=\columnwidth]{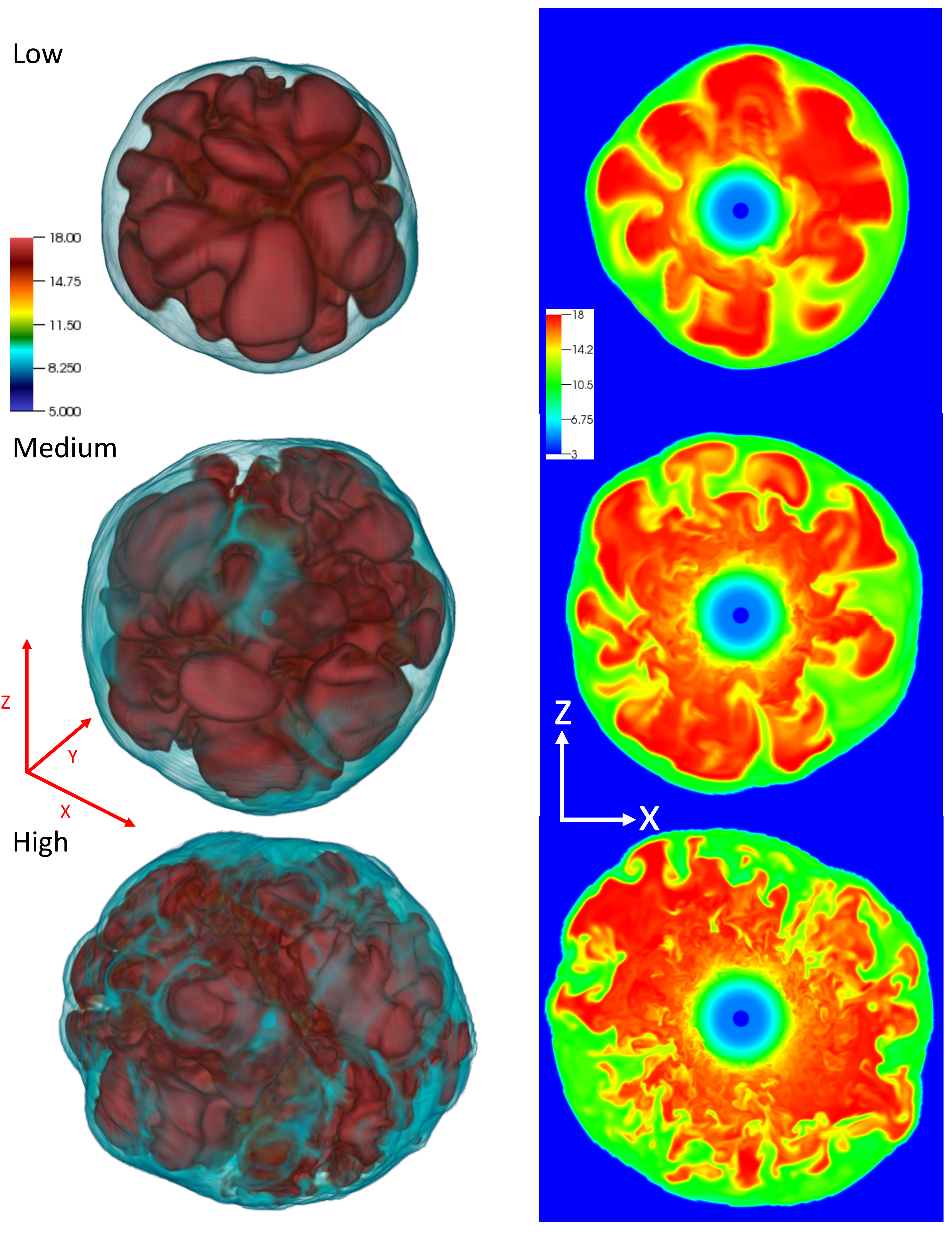}
    \caption{The same as Figure \ref{3D_ent1}, but at 200 milliseconds after bounce. Here, the 
spatial scale is $2\times{270}$ km.} 
    \label{3D_ent2}
  \end{minipage}
\end{figure*}

Figures \ref{3D_ent1} and \ref{3D_ent2} depict stills at 100 and 200 milliseconds after bounce
of the entropy field of the core for models 3DL, 3DM, and 3DH.  The improving resolution from top to bottom
is clearly manifest, with the smaller scales coming into sharper focus as we progress from model 3DL to model 3DH.
The resolution of model 3DL is at or near the standard resolution employed in the literature. The 
resolution of model 3DM has to date been our default resolution \citep{2019MNRAS.482..351V,2019MNRAS.485.3153B}. 
As noted in \S\ref{sec:methodandmodel}, each model employs a factor of two better resolution in both 
$\theta$ and $\phi$ as we step up from model 3DL, through model 3DM, to model 3DH. With the slight 
expansion of the shock radius, the 200-ms stills capture the onset of the explosion of models 3DM and 3DH.

\subsection{Neutrino Emission and Heating}
\label{sec:neutrinoheating}

\begin{figure*}
  \begin{minipage}{0.75\hsize}
	\includegraphics[width=\columnwidth]{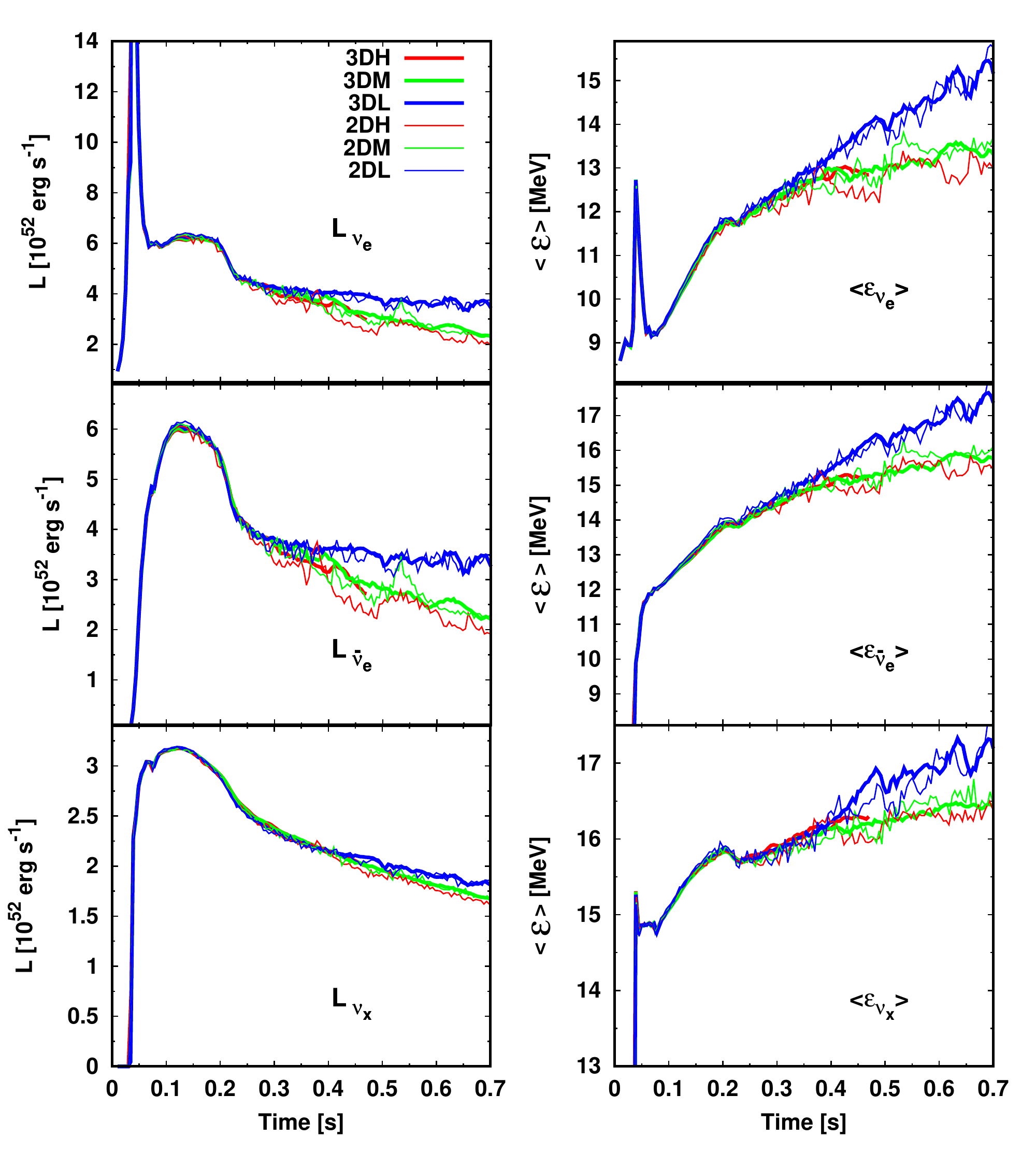}
    \caption{{\bf Left:} The total neutrino luminosity (in units of $10^{52}$ ergs s$^{-1}$) for the 
three neutrino species (the last for each of the $\nu_{\mu}$, $\bar{\nu}_{\mu}$, $\nu_{\tau}$, and 
$\bar{\nu}_{\tau}$ species separately) versus time after bounce (in seconds) for the three resolutions
(L, M, and H) chosen for this study.  The thin lines are for the 2D models and the thick lines 
are for the 3D models.  {\bf Right:} The same as the left, but for the corresponding mean 
neutrino energies (in MeV). Note that the 2D models show more temporal variation than the 
3D models, due in part to the artificial axial sloshing motions seen in generic 2D simulations.
Such motions are rarely seen in 3D simulations and are in part artifacts of the 2D constraint. 
As the figures demonstrate, the neutrino emission characteristics in 2D and 3D are quite similar.}
    \label{figTevoLumi}
  \end{minipage}
\end{figure*}

\begin{figure*}
  \begin{minipage}{0.6\hsize}
        \includegraphics[width=\columnwidth]{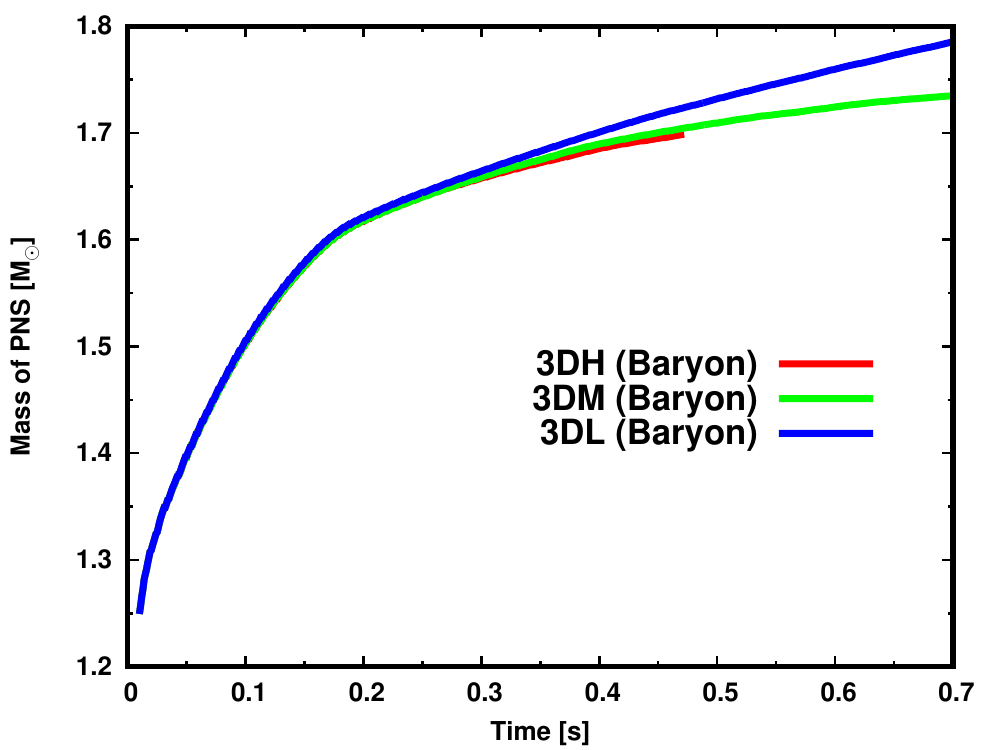}
    \caption{The evolution of the PNS baryon mass (in M$_{\odot}$) with time after bounce (in seconds)
for the three 3D models of this study.  Note that the accumulations of mass in the PNS for models 3DM 
and 3DH separate from that for model 3DL as a result of the latter's failure to explode.}
    \label{graph_timetrajectories_PNSmass}
  \end{minipage}
\end{figure*}

Figure \ref{figTevoLumi} compares the solid-angle-summed total neutrino luminosities and the 
mean neutrino energies versus time after bounce for the different neutrino species and for
the three 3D models (thick lines).  We also show for comparison the corresponding values for the 2D models
(thin lines).  As noted previously, the 2D model neutrino emissions fluctuate more than the 3D models, but are 
otherwise quite similar. Importantly, before explosion models 3DL, 3DM, and 3DH behave strikingly 
similarly and give little hint of the manifest bifurcation in late-time behavior.  After the explosion
of models 3DM and 3DH, the associated gradual cessation of accretion translates into a decrease
in the accretion component of their neutrino luminosities.  However, the continuing accretion of 
model 3DL, due to the fact it does not explode, both increases its core mass at a more 
rapid rate (see Figure \ref{graph_timetrajectories_PNSmass}) and maintains at a higher 
level accretion's contribution to the neutrino luminosities. This is the origin of the 
separation in the luminosity and average neutrino energy curves depicted in Figures 
\ref{figTevoLumi} after the explosions of models 3DM and 3DH. \citet{2018MNRAS.480.4710S} discuss
the possible observational discriminants of explosion versus no-explosion in underground 
terrestrial neutrino detectors.

\begin{figure*}
  \begin{minipage}{0.6\hsize}
	\includegraphics[width=\columnwidth]{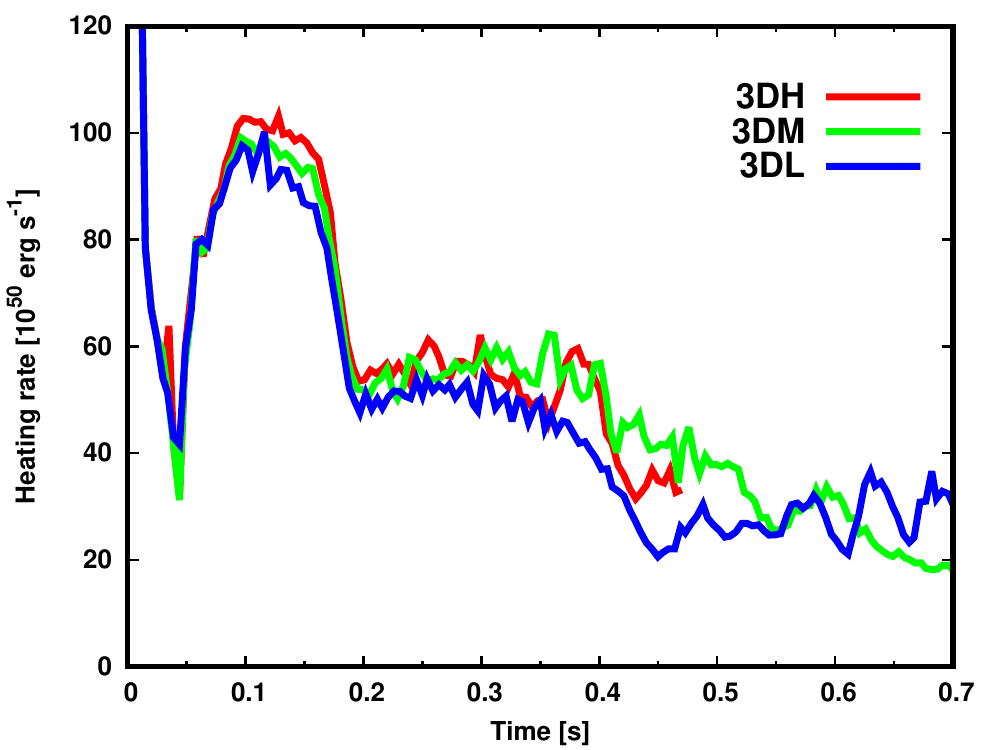}
        \includegraphics[width=\columnwidth]{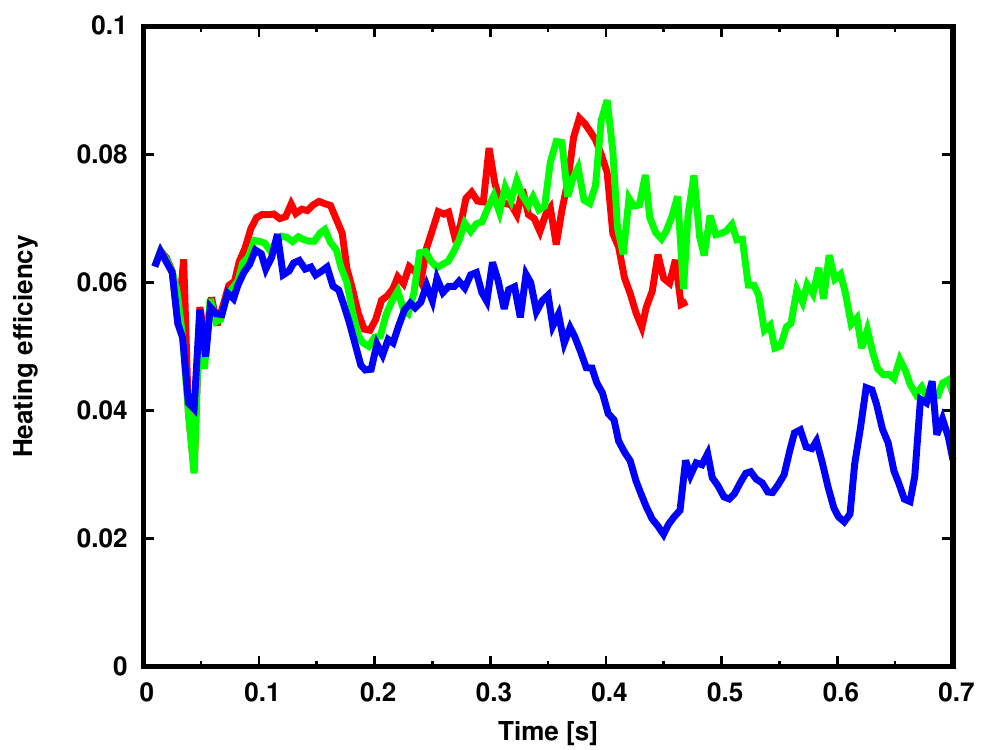}
    \caption{{\bf Top:} The neutrino heating rate in the gain region (in units of $10^{51}$ ergs s$^{-1}$) 
versus time after bounce (in seconds) for models 3DL (blue), 3DM (green), and 3DH (red). Heating due to
inelastic neutrino-electron and neutrino-nucleon scattering, though included in the simulations, 
is not included in this rate. {\bf Bottom:} Same as at the top, but for the corresponding 
heating efficiency. The latter is defined as the ratio of the heating rate to the sum of 
the total $\nu_{\rm e}$ and $\bar{\nu}_{\rm e}$ luminosities.  A 5-millisecond boxcar average 
has been applied. Note that after $\sim$100 ms and before $\sim$200 ms the heating rate 
in the gain region increases with resolution. See text for a discussion.}
    \label{graph_heating}
  \end{minipage}
\end{figure*}

In Figure \ref{graph_heating}, however, we start to see slight differences in the neutrino sector
between the models. This figure renders the heating rate in the gain region \citep{1985ApJ...295...14B}
due to neutrino absorption as a function of time after bounce for the three 3D models and 
the associated heating efficiency.  The latter is defined as the ratio between this heating rate
and the sum of the $\nu_{\rm e}$ and $\bar{\nu}_{\rm e}$ luminosities and is, hence, a measure of the 
``optical depth" to neutrino absorption in the gain region.  We see that after 
$\sim$100 ms the more resolved models have slightly higher neutrino heating rates and efficiencies.
This parallels the slightly larger mean shock radii seen in Figure \ref{graph_timetrajectories_shockradii} 
for the higher resolutions.  This trend is in part an explanation of the resolution 
dependence in the explodability we find. As we articulate in \S\ref{turbulence}, the heightened turbulent 
pressures behind the shock in the gain region are likely responsible for the larger pre-explosion shock radii 
(Figure \ref{graph_timetrajectories_shockradii}) and larger gain region.  
In fact, we witness that the mean radius of the stalled shock before explosion for our highest-resolution model (3DH) is 
$\sim$10$-$20 km greater than that for our lowest-resolution model (3DL), and that this difference correlates 
with a higher Reynolds stress behind the shock.
Since the neutrino luminosities and neutrino energies are not much affected by resolution (see Figure \ref{figTevoLumi}), they
are not the explanation for the slight augmentation in the heating rate with resolution.  Rather, the larger 
rate of neutrino energy deposition behind the shock in the gain region is a consequence
of its larger geometric size and the associated larger optical depth to absorption, as the bottom 
panel of Figure \ref{graph_heating} demonstrates. Therefore, turbulence not only pushes the shock radius
further out, but it bootstraps a concomitant increase in the deposited neutrino power.  Both effects
together lie at the core of the resolution dependence we witness. We now turn to a discussion 
of the turbulence and its resolution dependence.

\subsection{Turbulence}
\label{turbulence}

Neutrino-driven convection behind the stalled shock wave has been studied for decades
\citep{herant1994,1995ApJ...450..830B,2002ApJ...574L..65F,Murphy:2008dw,2013ApJ...770...66H,murphy:13} and
has been shown to be an essential factor in igniting the supernova within the neutrino-driven paradigm
\citep{1995ApJ...450..830B,janka_96,Murphy:2008dw,murphy:13,dolence:13}.  The Rayleigh-Taylor-like 
instability that naturally arises due to neutrino heating from below and that transitions into 
non-linear turbulence results in larger stalled shock radii \citep{burrows2013}\footnote{The 
standing-accretion-shock-instability (SASI) \citep{blondin2003} may also play a role.}. 
In this regard, the major relevant aspect of neutrino-driven turbulence seems to be the generation of 
Reynolds stress associated with its chaotic motions\footnote{The increase in the average dwell time 
in the gain region of a heated Lagrangian mass element may also be a minor factor 
\citep{1995ApJ...450..830B,Murphy:2008dw,2012ApJ...749...98T}.}. Channeling accretion energy in part into turbulence, 
instead of into thermal energy, provides stress/pressure more efficiently for a given amount of energy, 
since the effective $\gamma$ of the turbulent motions is $\sim$2, and not the $\sim{4/3}$ of gas
\citep{1995ApJ...450..830B,murphy:13,2016ApJ...820...76R}.  

This elevated total post-shock 
stress pushes the shock wave out by $\sim$10's of kilometers relative to spherical models for
which overturn instabilities are suppressed, thereby decreasing the pre-shock ram pressure 
experienced and placing matter that resides just behind the shock shallower in the gravitational 
potential well.  Hence, by going to multi-D and allowing the core to achieve a lower 
free energy through the agency of hydrodynamic instability, explosion can be facilitated 
in theoretical models.  It is thought that Nature may be doing the same.  

Hydrodynamic turbulence per se in core-collapse and proto-neutron stars has received 
some theoretical attention, with the perennial focus being on the degree of convergence of the flow
and the resolution necessary to capture the inertial range \citep{2014ApJ...785..123C,2015ApJ...808...70A,
2015ComAC...2....7R,2016ApJ...820...76R,2019arXiv190401699M}.  The consensus has been that spatial
resolutions beyond what is currently computationally feasible are necessary to achieve 
the latter\footnote{It has been suggested \citep{1988PhR...163...51B} that neutrino viscosity itself could truncate 
the inertial range in some parts of the core with low Reynolds numbers, perhaps $\sim$100's to $\sim$1000's, 
and that the relevant dissipative scale is much larger than that determined by the microscopic viscosity of 
matter. Codes such as F{\sc{ornax}} that incorporate radiation stress and full velocity-dependent transport
in a multi-D context naturally contain this physics and can be used to explore this phenomenon.  
We leave to future work an investigation of this interesting possibility.}, but that 
the character of the flow on larger scales and the energy flux from large to small scales of the Kolmogorov
cascade can stabilize at the actual physical value with current codes and computers.  This has
been demonstrated in the past by the observation that at larger and intermediate scales on the grid
the energy density spectrum has stopped changing as the resolution has increased and stabilizes
at the classic five-thirds law (see, e.g., \citet{dolence:13,2015ApJ...808...70A,2016ApJ...820...76R}).  
So, without addressing the technical question of the resolution of turbulence in the full inertial range, 
we nevertheless explore with confidence in this paper the dependence upon resolution of 
state-of-the-art CCSN simulations and their outcomes. 

Due to the fact that in the CCSN problem matter advects through the turbulent region and that this region 
is not a closed box, the inauguration of instability and its progress depends upon perturbations advected
through the shock, the advection speed, and the size of the gain region \citep{2006ApJ...652.1436F,2015PASA...32....9F}.
Nevertheless, a qualitative analysis of the character of the turbulence can proceed using the standard
metrics.

The turbulent kinetic energy spectrum $E$($\kappa$ or $\ell$) (versus the wavenumber, $\kappa$,
or the spherical-harmonic index, $\ell$) helps us to assess the nature of turbulence behind the shock
in the CCSN context.  In the Kolmogorov scheme, as generalized by Pao \citep{pao_1965}, $E(\kappa)$, the spectrum
of the specific turbulent kinetic energy, is generally written as:
\begin{equation}
E(\kappa) = \alpha \epsilon^{2/3}\kappa^{-5/3} e^{-\frac{3}{2}\alpha(\eta\kappa)^\frac{4}{3}}, \,    
\label{kolmo}
\end{equation}
where $\eta$ is the dissipative scale, $\epsilon$ is the Kolmogorov dissipation rate, $\kappa$ is the 
spatial wavenumber of the specific kinetic energy spectrum, and $\alpha$ is an empirical dimensionless constant 
near $\sim$1.6.  The exponential term is an approximate way \citet{pao_1965} derived to truncate 
the otherwise $-5/3$ power-law cascade at the dissipative scale with a constant viscosity model 
that mimics real viscosity.  If it is assumed that ``numerical viscosity" behaves in this 
fashion, fitting eq. (\ref{kolmo}) to the numerical kinetic energy spectrum $E(\kappa)$ would 
provide both the effective dissipative scale (``$\eta$") and the effective numerical viscosity.  
The validity of this approach is contingent upon the specific code and is unlikely to be 
true in detail.  It is nevertheless a useful way to conceptualize the effective magnitude of both 
numerical viscosity and the numerical dissipative scale. The Reynolds number ($Re$) of the turbulence 
is approximately $(\frac{L}{\eta})^{4/3}$, where the $L$ is approximately the size of the convective 
zone (here, the width of the gain region).  If $\eta$ were the physical scale at which physical viscosity truncates the inertial 
range it could be quite small and $Re$ would be quite large.  However, in a numerical simulation
$\eta$ is likely some small multiple of the smallest linear grid scale and the effective $Re$ is small 
\citep{2015ComAC...2....7R,2016ApJ...820...76R}. Back-of-the-envelope Kolmogorov arguments would then yield an 
effective viscosity $\nu \sim \epsilon^{1/3} \eta^{4/3}$.  This could also be derived by dividing
the numerical Reynolds number $Re$ into the product of turbulent speed on the largest/driving scales
and $L$. However, in this paper we are mainly interested in the empirical trends of these 
quantities and the supernova dynamics with grid angular resolution and provide equation 
(\ref{kolmo}) merely for guidance and context. 

When using a spherical-polar grid, calculating $E(\kappa)$ in terms of the spherical-harmonic $\ell$
at a given radius $R$ is more convenient and there is a simple one-to-one mapping between $\ell$ 
and $\kappa$ $(\kappa^2 = \ell(\ell+1)/R^2)$.  Therefore, we use the formalism of \citet{2015ApJ...808...70A}.
Moreover, we calculate the spectrum of the kinetic energy density, $\cal{E}(\ell)$, as opposed to the spectrum
of the specific kinetic energy (eq. (\ref{kolmo})). The integral under the corresponding linear-linear 
curve provides the overall average of the kinetic-energy density.  

\begin{figure*}
  \begin{minipage}{0.5\hsize}
        \includegraphics[width=\columnwidth]{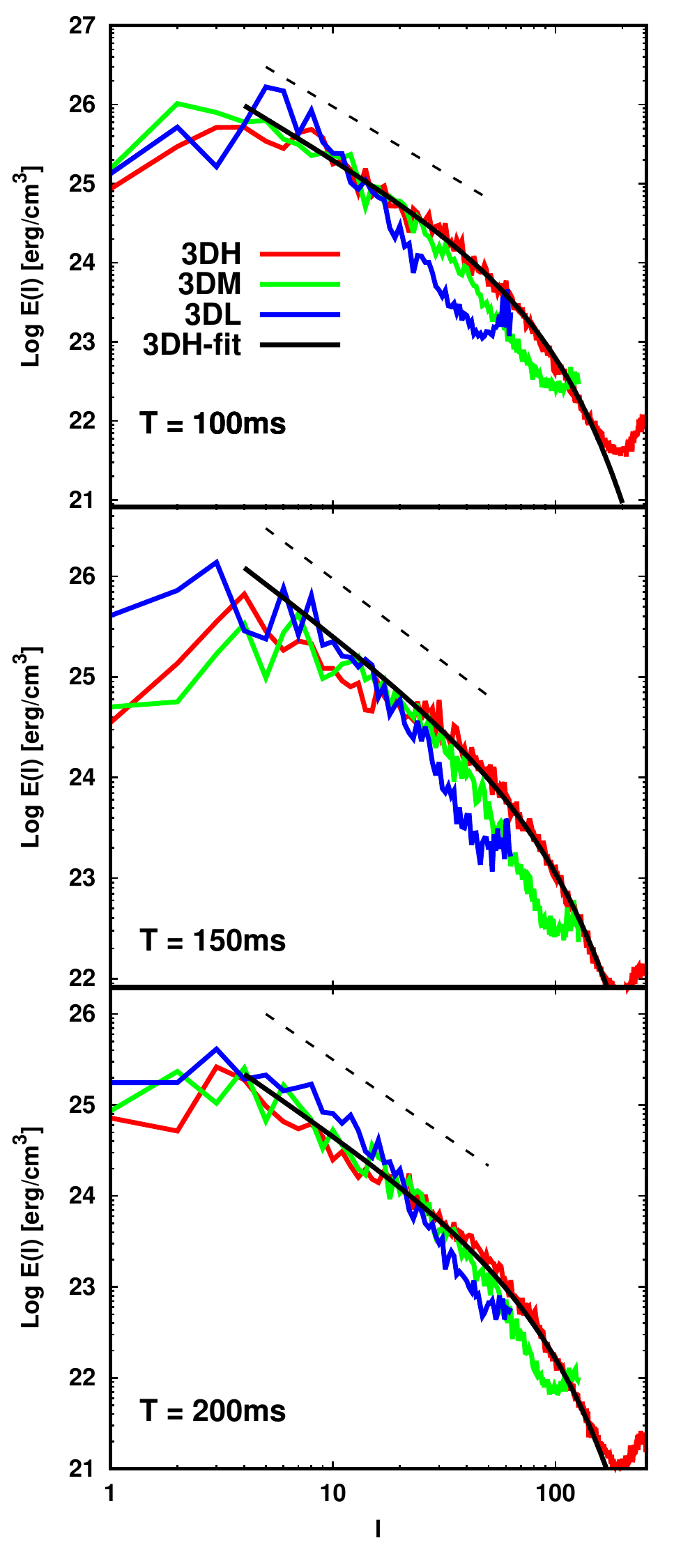}
    \caption{The power spectrum of the turbulent kinetic energy density in the transverse direction
versus spherical harmonic index $\ell$ at three different times post-bounce and for the three
models 3DL, 3DM, and 3DH with low, medium, and high resolution, respectively, as defined in the text. 
The region sampled and averaged is between the shock and the isodensity surface at 10$^{11}$ g cm$^{-3}$. Superposed
is a line tracing a $-\frac{5}{3}$ slope. The upturn at the highest values of $\ell$ is a numerical
artifact as we approach the grid scale at the highest values of $\ell$.  Also, included are 
Pao-hypothesis fits (black; eq. \ref{kolmo}) to the spectra of the highest-resolution model 3DH. See text for a discussion.
}
    \label{turb}
  \end{minipage}
\end{figure*}

\begin{figure*}
  \begin{minipage}{0.5\hsize}
        \includegraphics[width=\columnwidth]{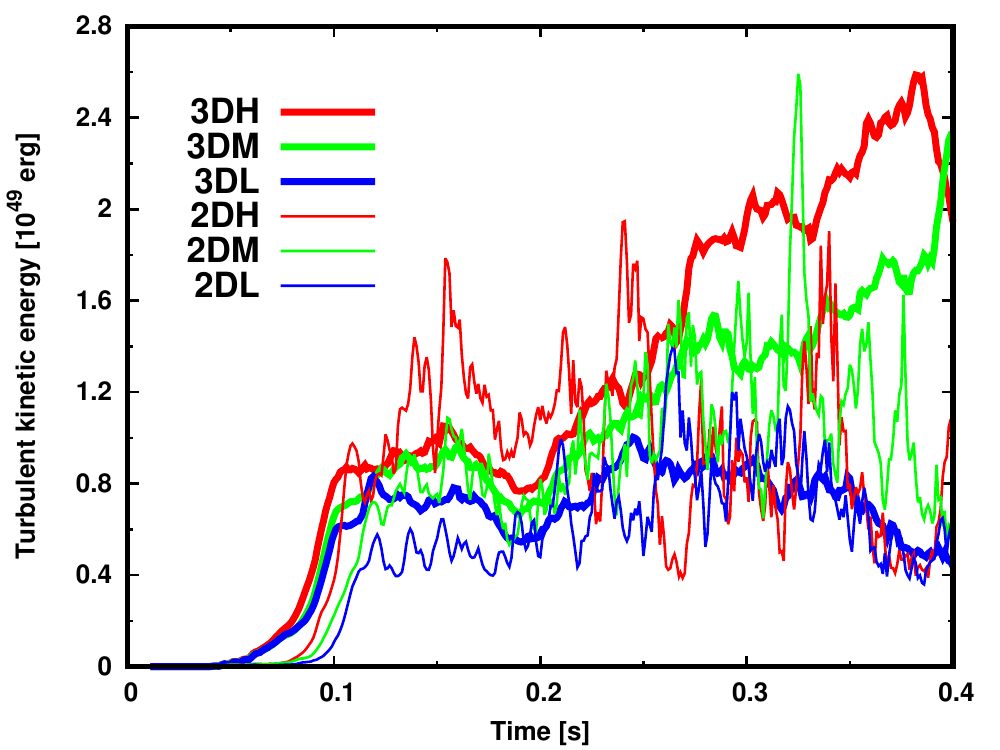}
    \caption{The transverse component of the turbulent kinetic energy (in units of 10$^{49}$ ergs)
in the gain region versus time (in seconds) after bounce. Its increase with resolution is clearly seen for both the
2D and 3D set of simulations. See text for a discussion.}
    \label{turbke}
  \end{minipage}
\end{figure*}


Figure \ref{turb} renders this spectrum for the transverse ($\theta$ and $\phi$) kinetic energy density
at three different times after bounce for the three models, 3DL (blue), 3DM (green), and 3DH (red).  
The dashed line depicts a $-\frac{5}{3}$ slope. We recall from Figure \ref{graph_timetrajectories_shockradii} 
that models 3DM and 3DH explode near $\sim$200 ms.  As Figure \ref{turb} shows, there is a modest difference between
the calculated spectra for model 3DL and those of the other two, particularly at early times. The inertial range at the 
early times for model 3DL is captured not at all, but at later times it is captured over less than a factor 
of ten in $\ell$.  Models 3DM and 3DH are more similar at all times and each follows the $-\frac{5}{3}$
law over a wider range, with model 3DH doing so over a bit more than an order of magnitude.  This is better
than that obtained by \citet{2015ApJ...808...70A} and \citet{2016ApJ...825....6S}, and better than
all but the highest resolution leakage or light-bulb models in \citet{2015ComAC...2....7R}, 
\citet{2016ApJ...820...76R}, and \citet{2019arXiv190401699M}. Note that we apply static mesh 
refinement (the dendritic grid) in $\theta$ only interior to the turbulent gain region, 
while it is applied in $\phi$ near the pole all along the z-axis.  For model 3DH, by eye the turnover 
at higher $\ell$s that reflects the transition to the numerical dissipative regime 
occurs near $\ell = 70-80$. The turnover $\ell$s for the 3DL and 3DM models 
are $\sim$40 and $\sim$60, respectively.  The $\frac{R}{\eta}$ for the 3DH model, derived from 
a formal fit to eq. (\ref{kolmo}) for model 3DH, ranges from 50 to 58.  This translates 
for that higher-resolution model into a delta in angle of 
$\sim$2-3$^{\circ}$\footnote{Recall that our best $\Delta\theta$ is 0.7$^{\circ}$.}. 
In the gain region of model 3DH, this encompasses $\sim$4 angular zones.  Hence, it is not 
unexpected that our numerical inertial range would not extend much further in $\ell$.

The implication of the trend with resolution in turnover in $\ell$ seen in Figure \ref{turb} is that 
the numerical viscosity is higher and the numerical Reynolds number is lower for lower 
resolution.  As Figure \ref{turb} implies, the ratio of the turnover $\ell$s and, hence of the turnover $\eta$s 
and dissipative scales, for models 3DL and 3DH is 2$-$4. This translates into an increase
in the effective numerical viscosity (if it scales as $\sim \epsilon^{1/3} \eta^{4/3}$) 
of a factor of 2.5$-$6 in going from model 3DH to 3DL.  Even if the exact scaling is not 
as given by Pao's formula (eq. \ref{kolmo}), this general argument clearly suggests there is greater ``drag" on 
the turbulent flow for lower angular resolution. We see the manifestations of this in the figures below.
Importantly, however, we find that the general level in the inertial range of the turbulence spectra for models
3DM and 3DH is the same, suggesting that the Kolmogorov
dissipation rate
$\epsilon$ is also the same for these 
models and that the cascade flux through $\kappa$ space has stabilized at a physical value.    
Moreover, the higher $\ell$ values at which model 3DH turns over are one indication that 
the transverse turbulent energy density for it is larger.  Figure \ref{turbke} depicts
the associated clear trend with resolution of the total transverse kinetic energy in the 
gain region.  
Note that we observe the same trend with resolution in 2D as in 3D, i.e., higher resolution results in more turbulent kinetic energy. Furthermore, the time-averaged summed turbulent kinetic energy is not much different between 2D and 3D, given the same lateral resolution. This is consistent with our claim that explodability is not much different between 2D and 3D
\footnote{With regards to Fig.~\ref{turbke}, the greater time variability in 2D is likely due to the artificial axial sloshing in 2D. As a result, the turbulent kinetic energy in 2D frequently exceeds that in 3D (see, e.g., $\lesssim 250$ms in the high resolution models). We also note that Figure~\ref{turbke} indicates that the turbulent kinetic energy of the highest resolution model in 2D is lower than that in 3D after shock revival. Note that we compute the turbulent energy in the region where the outer boundary is set to the minimum shock radius. At later times in 2D, this tends to be smaller than in 3D due, again, to the articifial axial sloshing. This results in lower turbulent kinetic energy than in 3D. Finally, it is important to mention that our conclusions are not perfectly consonant with some previous work (e.g., \citet{2015ApJ...799....5C,2016PASA...33...48M}), in which it is claimed that the strength of turbulence in 2D is larger than that in 3D. Currently, we have not identified the primary cause of this difference, which will only be revealed by making detailed group-to-group comparisons.}.


The anisotropic Reynolds stress tensor ($R_{ij}$) is:

\begin{equation}
R_{ij} = <\rho {v}^{\prime}_i{v}^{\prime}_j> \, ,
\end{equation}
where ${v}^{\prime}_i$ is the $i$th component of the turbulent velocity after subtraction
of the mean flow (taken here to be radial) and $\rho$ is the mass density. $<{\cdots}>$ is the 
angle
average.
For radial support, the radial component of the Reynolds stress ($R_{rr}$) is the most important 
and is generally larger than the corresponding value for the transverse $\theta{\theta}$ and
$\phi{\phi}$ components \citep{murphy:13}. $R_{ij}/P$, where $P$ is the gas pressure, is a metric of
the relative contribution of the Reynolds stress to the total stress/pressure, and, to within the $\gamma$ of the gas,
is the average of the square of the Mach number ($M$) of the turbulence.  These metrics are
useful gauges of the importance of turbulence in the mean flow dynamics \citep{2015MNRAS.448.2141M,2016ApJ...825....6S}.  

\begin{figure*}
  \begin{minipage}{0.7\hsize}
        \includegraphics[width=\columnwidth]{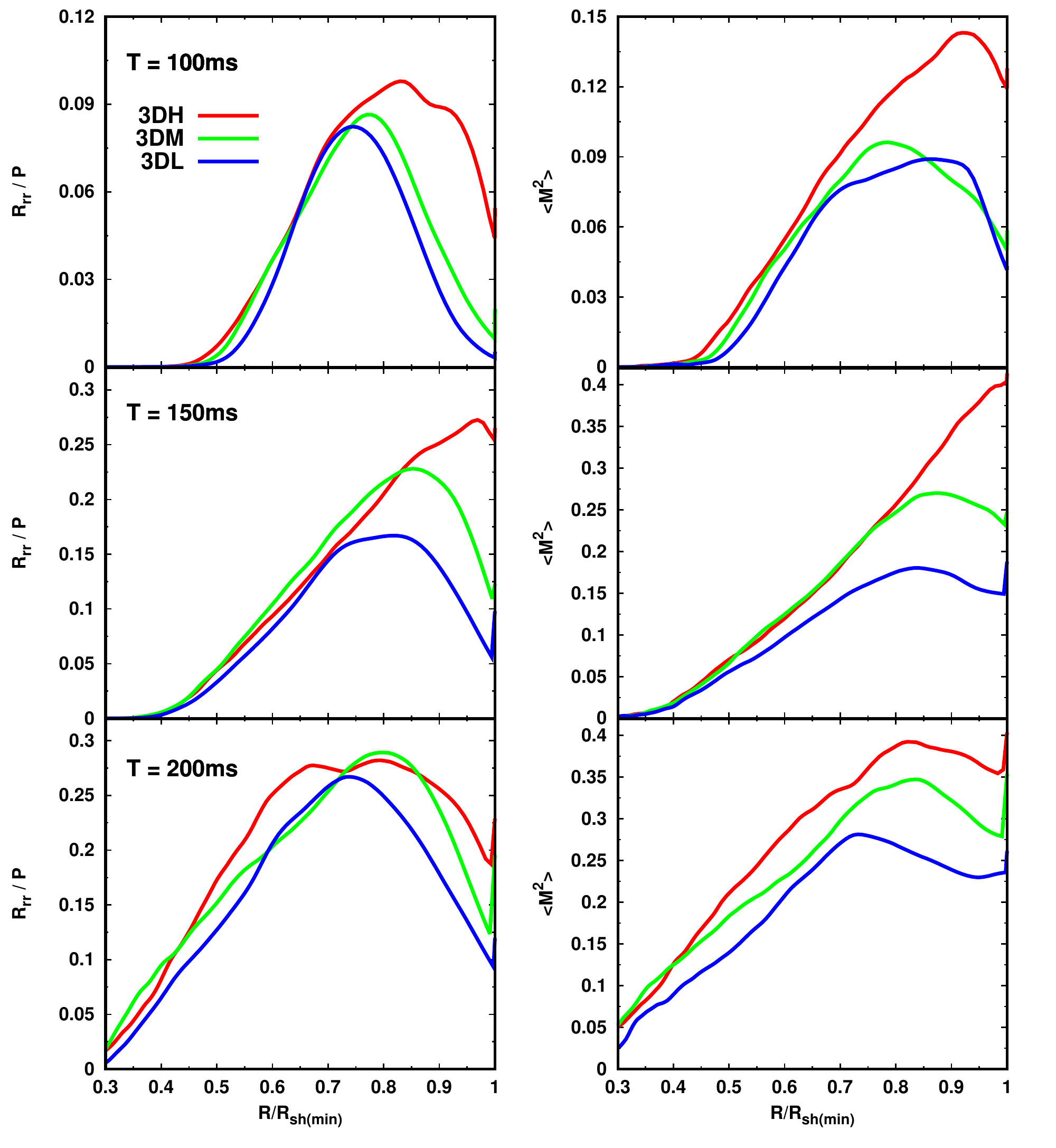}
    \caption{{\bf Left:} The ratio of the radial component of the turbulent stress ($R_{rr}$) and the gas pressure,
versus the radius, normalized to the instantaneous minimum shock radius, for the three models 3DL, 3DM, and 3DH,
for three different times after bounce (100, 150, and 200 ms). {\bf Right:} The average of the square of the 
total Mach number.  All three velocity components are included in this number, added in quadrature. 
Note that in general, but particularly near the shock wave, these metrics of the strength
of the turbulent stress increase with resolution. See text for a discussion.}
    \label{reynolds_r}
  \end{minipage}
\end{figure*}

\begin{figure*}
  \begin{minipage}{0.7\hsize}
        \includegraphics[width=\columnwidth]{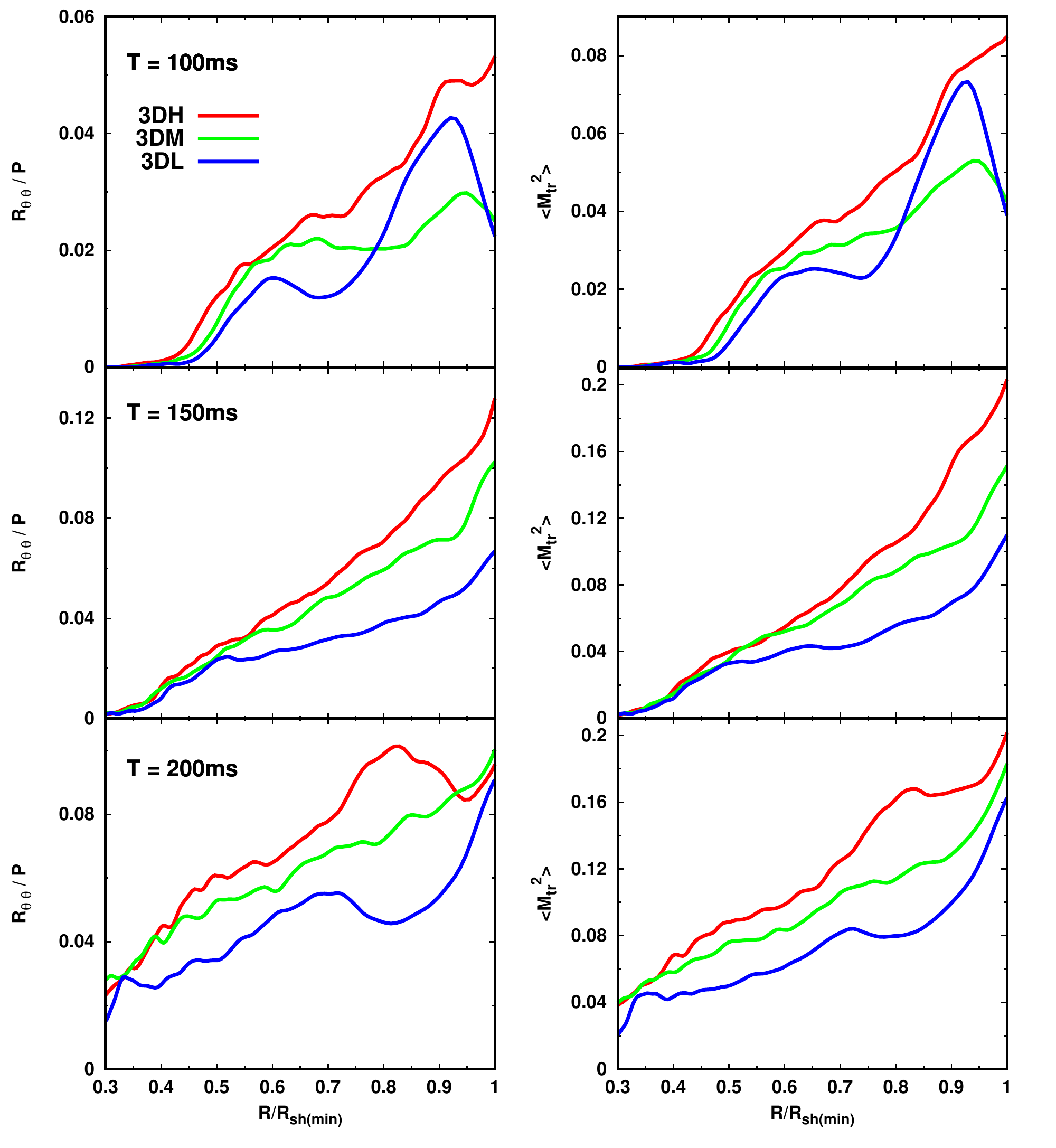}
    \caption{The same as Figure \ref{reynolds_r}, but for $R_{\theta\theta}$ and the average Mach 
number squared for only the $\theta$ and $\phi$ components of the velocity.  $R_{\phi\phi}$ is 
comparable to $R_{\theta\theta}$.}
    \label{reynolds_t}
  \end{minipage}
\end{figure*}

The left panel of Figure \ref{reynolds_r} portrays the radial profile of the radial stress 
divided by the gas pressure, $R_{rr}/P$, for the various 3D models and for post-shock times 
of 100, 150, and 200 milliseconds.  Its right panel portrays the associated average of the square 
of the turbulent Mach number ($<M^2>$).  The radius is normalized to the minimum shock radius for  
the given snapshot and the given model. Figure \ref{reynolds_t} provides the same quantities for $R_{\theta\theta}$
and the associated $<M^2>$.  To calculate $R_{rr}$, we subtract out the solid-angle-averaged 
radial speed at the given radius in the gain region. We do not need to 
subtract a mean in calculating $R_{\theta\theta}$. 

As demonstrated with Figures \ref{reynolds_r}
and \ref{reynolds_t}, the contribution of the turbulent stress to the total stress behind the shock 
in the gain region is generally larger for better resolved models.  As the earlier turnoff in $\ell$
found in Figure \ref{turb} suggests, numerical viscosity is largest for the least resolved
model.  As Figures \ref{reynolds_r} and \ref{reynolds_t} suggest, this translates into weaker 
turbulence for the least resolved models and stronger turbulence as angular resolution improves.
As stated in \S\ref{sec:neutrinoheating}, the greater turbulent stress pushes the stalled shock 
radius further out, which in turn results in greater neutrino power deposition in the gain region.
Together, and as indicated in Figure \ref{graph_timetrajectories_shockradii}, these effects make 
a better resolved model more explodable.  However, what resolution is required to achieve a converged
solution has yet to be determined. Nevertheless, that increasing resolution better supports
explosion is both a boost to the theoretical viability of the neutrino mechanism and a cautionary tale $-$
along with the microphysics, spatial resolution should be a more central focus of CCSN theory.

\section{Conclusions} \label{sec:conclusions} 

We have endeavored with this paper to explore the dependence upon spatial resolution of the dynamics 
and explodability of 3D core-collapse supernova simulations.  To this end, we enlisted our state-of-the-art
CCSN code F{\sc{ornax}} to the determine what differences would emerge when changing only the number
of angular bins in the $\theta$ and $\phi$ directions, all else kept exactly the same.  We used the same
19-M$_{\odot}$ progenitor, the same energy and radial binning, and the same 
microphysics and EOS. What we discovered was that our lowest resolution simulation (model 3DL) did not explode,
while when jumping progressively up in resolution by factors of two in each angular direction 
on our spherical-polar grid models then exploded, and exploded a tad more vigorously with increasing resolution.
This suggests that there can be a qualitative dependence upon spatial resolution of the outcome of CCSN 
simulations, but importantly that increasing the resolution (and presumably the accuracy of the calculations) 
may bring the models closer to Nature.  We have not, however, proven the latter assertion, but had
increases in the resolution inhibited explosion the viability of the neutrino mechanism of core-collapse supernova
explosions, at least as addressed with modern codes and implementations, might have been in doubt.
Rather, we find that for a given code and algorithm, and for a given suite of microphysics, there
may be a resolution below which a model that ``should" explode will not.   This also may suggest that striving 
for higher spatial resolution in 3D CCSN simulations may be as important as refining the comprehensive
suite of neutrino-matter interactions and identifying a viable EOS.  This conclusion may explain, at 
least in part, why some sophisticated published 3D models did not explode $-$ the resolution for 
those implementations might not have been adequate.  We stress that the required resolutions 
will be code and methodology dependent. This may also be the reason our published 3D 13-M$_{\odot}$ did not explode
\citep{2019MNRAS.485.3153B}, a possibility we are currently exploring.

What we find is that the critical aspect of higher spatial resolution is the adequate capturing of the physics 
of neutrino-driven turbulence, in particular its Reynolds stress.  The latter is an important factor in 
buoying the position of the stalled shock wave and later launching the supernova explosion.  The mean radius
of the stalled shock before explosion for our highest-resolution model was 10$-$20 km greater than that for our 
lowest-resolution model, and this difference correlated with a higher Reynolds stress behind the shock and a higher neutrino 
heating rate in the gain region. These differences are comparable to differences associated with those found 
when incorporating inelastic energy transfers, adding many-body corrections to neutrino-nucleon scattering 
rates, or embedding modest progenitor perturbations or rotation \citep{2018MNRAS.477.3091V,2018SSRv..214...33B}. 

Supernova theory is entering an exciting stage, wherein many sophisticated 3D models can now be executed
in routine fashion and at a respectable cadence to investigate the full range of topics associated with
core-collapse supernova explosions.  These topics include, among others, explosion energies, residual neutron star masses, 
pulsar kicks, massive star nucleosynthesis, black hole formation, and the morphology of supernova explosions 
and remnants. However, there are still limitations to current codes and studies.  Multi-angle, 
3D, long-duration simulations are still not within reach on current machines.  A fully
consistent implementation of general relativity with relativistic transfer/transport has yet to be fielded.
There remain issues of a quantitative character concerning the neutrino-matter interaction rates.
The equation of state of hot, lepton-rich nuclear matter is
a perennial concern, though laboratory and neutron-star constraints have been improving. And finally, the
massive star progenitor models inherited by CCSN modelers are still in flux. Even the detailed mapping 
between progenitor mass and density/temperature/composition/perturbation structure at the moment of collapse
is not settled, and we do not know the final rotation rate of the evolved cores of massive stars.  Despite 
all these concerns, there has been significant progress towards realizing the goal of a definitive 
understanding of the mechanism and nature of supernova explosions.  With the advent of modern 3D codes, 
advances in physical understanding, and the emergence of capable supercomputers,
supernova theory seems well poised to embark upon its next productive phase.

\section*{Acknowledgements}

The authors are grateful for ongoing contributions to
this effort by Josh Dolence and Aaron Skinner.
We also acknowledge Evan O'Connor regarding the equation of state,
Gabriel Mart\'inez-Pinedo concerning electron capture on heavy nuclei,
Tug Sukhbold and Stan Woosley for providing details concerning the
initial models, Todd Thompson regarding inelastic scattering,
and Andrina Nicola for help in computing the turbulent spectrum.
We acknowledge support
from the U.S. Department of Energy Office of Science and the Office
of Advanced Scientific Computing Research via the Scientific Discovery
through Advanced Computing (SciDAC4) program and Grant DE-SC0018297
(subaward 00009650). In addition, we gratefully acknowledge support
from the U.S. NSF under Grants AST-1714267 and PHY-1144374 (the latter
via the Max-Planck/Princeton Center (MPPC) for Plasma Physics). DR
cites partial support as a Frank and Peggy Taplin Fellow at
the Institute for Advanced Study. An award of computer time was provided 
by the INCITE program. That research used resources of the
Argonne Leadership Computing Facility, which is a DOE Office of Science 
User Facility supported under Contract DE-AC02-06CH11357. In addition, this overall research 
project is part of the Blue Waters sustained-petascale computing project,
which is supported by the National Science Foundation (awards OCI-0725070
and ACI-1238993) and the state of Illinois. Blue Waters is a joint effort
of the University of Illinois at Urbana-Champaign and its National Center
for Supercomputing Applications. This general project is also part of
the ``Three-Dimensional Simulations of Core-Collapse Supernovae" PRAC
allocation support by the National Science Foundation (under award \#OAC-1809073).
Moreover, access under the local award \#TG-AST170045
to the resource Stampede2 in the Extreme Science and Engineering Discovery
Environment (XSEDE), which is supported by National Science Foundation grant
number ACI-1548562, was crucial to the completion of this work.  Finally,
the authors employed computational resources provided by the TIGRESS high
performance computer center at Princeton University, which is jointly
supported by the Princeton Institute for Computational Science and
Engineering (PICSciE) and the Princeton University Office of Information
Technology, and acknowledge our continuing allocation at the National
Energy Research Scientific Computing Center (NERSC), which is
supported by the Office of Science of the US Department of Energy
(DOE) under contract DE-AC03-76SF00098.





\bibliographystyle{mnras}
\bibliography{bibfile}







\bsp	
\label{lastpage}
\end{document}